\documentclass{article}[9pt]
\usepackage{spconf,amsmath,graphicx}

\usepackage{amsmath}
\usepackage{amssymb}
\usepackage{amsfonts}
\usepackage{graphicx}
\usepackage{enumerate,color,graphicx,fancybox,pifont,epsf,epsfig,subfigure,amsmath,amssymb,psfrag}
\usepackage{algorithm}
\usepackage{algorithmic}
\usepackage{cite}

\newcommand{\secref}[1]{Section~\ref{#1}}

\newcommand{\figref}[1]{Figure~\ref{#1}}

\newcommand{\algref}[1]{Algorithm~\ref{#1}}

\newcommand{\qed}{\nobreak \ifvmode \relax \else
      \ifdim\lastskip<1.5em \hskip-\lastskip
      \hskip1.5em plus0em minus0.5em \fi \nobreak
      \vrule height0.75em width0.5em depth0.25em\fi}

\title{Channel--Optimized Vector Quantizer Design for \\ Compressed Sensing Measurements}
\name{Amirpasha Shirazinia, Saikat Chatterjee, Mikael Skoglund}
\address{Communication Theory Lab, ACCESS Linnaeus Centre \\
KTH Royal Institute of Technology, Stockholm, Sweden\\
{\small \tt Email: amishi@kth.se, sach@kth.se, skoglund@kth.se}}

\begin{document}
\ninept
\maketitle

\begin{abstract}
We consider vector-quantized (VQ) transmission of compressed sensing (CS) measurements over noisy channels. Adopting mean-square error (MSE) criterion to measure the distortion between a sparse vector and its reconstruction, we derive channel-optimized quantization principles for encoding CS measurement vector and reconstructing sparse source vector. The resulting necessary optimal conditions are used to develop an algorithm for training channel-optimized vector quantization (COVQ) of CS measurements by taking the end-to-end distortion measure into account.
\end{abstract}
\begin{keywords}
Channel-optimized vector quantizer, compressed sensing, sparsity, channel, mean-square error
\end{keywords}
\vspace{-0.15cm}
%%%%%%%%%%%%%%%%%%%%%%%%%%%%%%%%%%%%%%%%%%%%%%%%%%%%%%%%%%%%%%%%%%%%%%%%%%%%%%
\section{Introduction} \label{sec:intro}
%%%%%%%%%%%%%%%%%%%%%%%%%%%%%%%%%%%%%%%%%%%%%%%%%%%%%%%%%%%%%%%%%%%%%%%%%%%%%%
\vspace{-0.1cm}
Compressed sensing (CS) \cite{08:Candes} is a tool to retrieve a high-dimensional (nearly) sparse signal vector from low-dimensional measurement vector. In practical applications, CS measurements are required to be quantized into a finite-resolution representation, and then communicated to a destination point through a noisy channel for sparse signal reconstruction. Quantization and transmission errors are challenges imposed by practical systems, therefore, a robust approach against these imperfections is a fundamental design requirement which is taken in this paper.

Recently, significant research interest has been devoted to analysis and design of quantized transmission of CS measurements, and a wide range of interesting problems has been formulated. The extension of CS reconstruction methods in order to degrade the effect of quantization error has been considered in \cite{06:Candes2,10:Sinan,10:Zymnis,11:Dai,11:Jacques,12:Yan}, while in \cite{09:Sun,12:Boufounos,11:Kamilov,12:Pasha1,13:Pasha1}, the design of quantizers under fixed CS reconstruction methods has been studied. Further, theoretical bounds on distortions caused by CS reconstruction and quantization have been derived in \cite{08:Goyal,11:Dai,12:Laska}.

All of the aforementioned works are dedicated to pure source coding of CS measurements in which quantized transmission is assumed to be error-free. In practice, communication channel errors are often inevitable, and need to be considered through a quantizer design. To the best of our knowledge, quantized transmission of CS measurements over noisy channels has not been addressed. Therefore, in this paper, we consider a linear CS system where a random sparse source is compressed, and the resulting noisy CS measurements are quantized, transmitted over a discrete memoryless channel (DMC), and finally reconstructed at a receiving-end. The main contributions of the paper are twofold: first, we aim to find necessary optimal quantizer encoding and decoding principles with respect to minimizing average end-to-end \textit{mean-square error} (MSE) in the presence of channel imperfections. Second, we use the necessary optimal conditions, and propose a training algorithm in order to design a channel-optimized vector quantizer (COVQ) for CS by taking the end-to-end distortion criterion into consideration. 
We examine the performance of our proposed algorithm through simulation, and show its gain by comparing it vis-a-vis existing schemes for quantization of CS measurement vector.

\emph{Notations:} scalar random variables (RV's) will be denoted by upper-case letters while their instants will be denoted by the respective lower-case letters. Random vectors will be represented by boldface characters. 
Further, a set is shown by a calligraphic character and its cardinality by $|\cdot|$. We will also denote the transpose of a vector by $(\cdot)^T$.  We will use~$\mathbb{E}[\cdot]$ to denote the expectation operator. The $\ell_p$-norm ($p \geq 0$) of a vector will be denoted by $\|\cdot\|_p$.

\vspace{-0.4cm}
%%%%%%%%%%%%%%%%%%%%%%%%%%%%%%%%%%%%%%%%%%%%%%%%%%%%%%%%%%%%%%%%%%%%%%%%
\section{System Description} \label{sec:system}
\begin{figure}
  \begin{center}
  \psfrag{x}[][][0.75]{$\mathbf{X}$}
  \psfrag{A}[][][1]{$\mathbf{\Phi}$}
  \psfrag{y}[][][0.75]{$\mathbf{Y}$}
  \psfrag{w}[][][0.75]{$\mathbf{W}$}
  \psfrag{Q}[][][0.85]{$\textsf{E}$}
  \psfrag{C}[][][1]{$\mathcal{C}$}
  \psfrag{i}[][][0.75]{$I$}
  \psfrag{j}[][][0.75]{$J$}
  \psfrag{DMC}[][][0.8]{$P(j|i)$}
  \psfrag{Channel}[][][0.75]{Channel}
  \psfrag{quant}[][][0.75]{Quantizer}
  \psfrag{Enc}[][][0.75]{encoder}
  \psfrag{CS Enc}[][][0.75]{CS encoder}
  \psfrag{Dec}[][][0.75]{Decoder}
  \psfrag{Dec}[][][0.85]{Decoder}
  \psfrag{D}[][][0.85]{$\textsf{D}$}
  \psfrag{x_h}[][][0.75]{$\widehat{\mathbf{X}}$}
  \includegraphics[width=8.5cm]{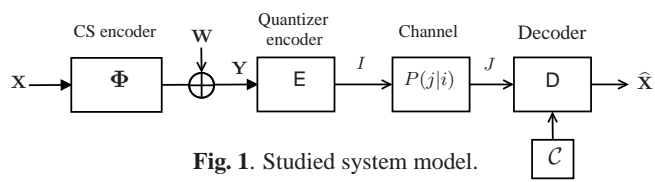}\\
  \vspace{-0.85cm}
  \caption{Studied system model.}\label{fig:diagram_CS}
  \vspace{-0.75cm}
  \end{center}
\end{figure}

%%%%%%%%%%%%%%%%%%%%%%%%%%%%%%%%%%%%%%%%%%%%%%%%%%%%%%%%%%%%%%%%%%%%%%%%
\vspace{-0.1cm}
In this section, we give an account for the basic assumptions and models made about the studied system depicted in \figref{fig:diagram_CS}.

\vspace{-0.3cm}
%%%%%%%%%%%%%%%%%%%%%%%%%%%%%%%%%%%%%%%%%%%%%%%%%%%%%%%%%%%%%%%%%%%%%%%%
\subsection{CS Framework} \label{subsec:CS}
%%%%%%%%%%%%%%%%%%%%%%%%%%%%%%%%%%%%%%%%%%%%%%%%%%%%%%%%%%%%%%%%%%%%%%%%
\vspace{-0.1cm}
Formally, we let a random sparse vector (where most coefficients are likely zero) $\mathbf{X} \in \mathbb{R}^M$ be linearly encoded using a known sensing matrix $\mathbf{\Phi} \in \mathbb{R}^{N \times M}$ ($N < M$) which results in an under-determined set of linear measurements, possibly perturbed by noise, i.e.,
\begin{equation} \label{eq:measurement}
    \mathbf{Y = \Phi X + W},
\end{equation}
where $\mathbf{Y} \in \mathbb{R}^N$ and $\mathbf{W} \in \mathbb{R}^N$ denote the measurement and the additive measurement noise vectors, respectively. We let $\mathbf{X}$ be an exact $K$-sparse vector, i.e., it has exactly $K$ ($K < N$) non-zero coefficients, where the location of non-zero's are uniformly drawn from all ${M \choose K}$ possibilities, and these components are identically and independently distributed (i.i.d.) standard Gaussian RV's. We also assume that the sparsity level $K$ is known in advance. We define the support set of the sparse vector $\mathbf{X} = [X_1,\ldots,X_M]^T$ as $\mathcal{S} \triangleq \{m : X_m \neq 0 \} \subset \{1,\ldots,M\}$ with $|\mathcal{S}| = \|\mathbf{X}\|_0 = K$. 
The elements of the sensing matrix $\Phi_{ij}$  are i.i.d. RV's drawn from a Gaussian distribution (i.e., $\Phi_{ij} \sim \mathcal{N} (0,1/N)$), where the columns of $\mathbf{\Phi}$ are normalized to unit-norm. Note that once $\mathbf{\Phi}$ is generated, it remains fixed and is made known to decoder. In order to reconstruct an unknown sparse source $\mathbf{X}$ from a noisy under-sampled measurement vector $\mathbf{Y}$, several reconstruction methods have been developed based on convex relaxation optimization problem (see e.g.~\cite{06:Candes2,07:Candes,08:Shihao}), iterative greedy search algorithms (see e.g. \cite{07:Tropp,08:Blumensath,09:Dai,12:Saikat}) and Bayesian estimation approaches (see e.g. \cite{07:Larsson,09:Elad,10:Protter}). 
\vspace{-0.3cm}
%%%%%%%%%%%%%%%%%%%%%%%%%%%%%%%%%%%%%%%%%%%%%%%%%%%%%%%%%%%%%%%%%%%%%%%%
\subsection{Vector Quantization over Noisy Channels} \label{subsec:VQ}
%%%%%%%%%%%%%%%%%%%%%%%%%%%%%%%%%%%%%%%%%%%%%%%%%%%%%%%%%%%%%%%%%%%%%%%%
\vspace{-0.1cm}
We consider the problem of vector quantization (VQ) of the linear noisy CS measurement vector $\mathbf{Y}$ over a discrete memoryless channel (DMC). We assume that the total bit budget allocated for quantization is fixed at $B$ bits per dimension of the source vector. Then, given the noisy measurement vector $\mathbf{Y}$, a VQ encoder is defined by a mapping $\textsf{E}: \mathbb{R}^N \rightarrow \mathcal{I}$, where $\mathcal{I}$ is a finite index set defined as $\mathcal{I} \triangleq \{0,1,\ldots,2^{B}-1\}$ with $|\mathcal{I}| \triangleq L=2^{B}$. Denoting the quantized index by $I$, the encoder works according to $\mathbf{Y} \in \mathcal{R}_i \Rightarrow I = i$, where the sets $\{\mathcal{R}_i\}_{i=0}^{L-1}$ are (possibly disconnected) encoder regions and $\bigcup_{i=0}^{L-1} \mathcal{R}_{i} = \mathbb{R}^N$ such that when $\mathbf{Y} \in \mathcal{R}_i$ the encoder outputs the index $\textsf{E}(\mathbf{Y}) = i \in \mathcal{I}$. 

Next, we consider a DMC which accepts the encoded index $i$, and outputs a noisy symbol, denoted by $j$. The channel is defined by a random mapping $\mathcal{I} \rightarrow  \mathcal{I}$ characterized by transition probabilities
\begin{equation} \label{eq:channel trans}
P(j | i) \triangleq \textrm{Pr}(J = j | I = i), \hspace{0.15cm} i,j \in \mathcal{I},
\end{equation}
which indicates the probability that index $j$ is received given that the input index to the channel was $i$. We assume that the transmitted index $i$ and the received index $j$ share the same index set $\mathcal{I}$, and the channel transition probabilities \eqref{eq:channel trans} are known in advance. Given the received index $j$, a combined VQ/CS decoder is characterized by a mapping $\textsf{D}: \mathcal{I} \rightarrow \mathcal{C}$ where $\mathcal{C}$ is a finite discrete \textit{codebook} set containing all reproduction \textit{codevectors} $\{\mathbf{c}_{j} \in \mathbb{R}^M\}_{j=0}^{L-1}$. Then, the decoder's functionality is described by a look-up table; $J = j \Rightarrow \widehat{\mathbf{X}} = \mathbf{c}_j$ such that when the received index from the channel is $j$, then the decoder outputs the codevector $\textsf{D}(j)=\mathbf{c}_j \in \mathcal{C}$.

It is important to design an encoder-decoder pair in order to minimize a distortion measure which reflects the need of the receiving-end user. Therefore, we quantify the performance of our studied system in \figref{fig:diagram_CS} by the end-to-end MSE defined as
\begin{equation} \label{eq:e2e dist}
    D \triangleq \mathbb{E}[\|\mathbf{X - \widehat{X}}\|_2^2].
\end{equation}
Note that the end-to-end MSE depends on CS reconstruction errors, quantization errors as well as channel errors, and our goal is to design an encoder/decoder pair which is robust against all three kinds of error.

\vspace{-0.3cm}
%%%%%%%%%%%%%%%%%%%%%%%%%%%%%%%%%%%%%%%%%%%%%%%%%%%%%%%%%%%%%%%%%%%%%%%%%%%%%%%%%%%%%%%
\section{Design Problem} \label{sec:conditions}
%%%%%%%%%%%%%%%%%%%%%%%%%%%%%%%%%%%%%%%%%%%%%%%%%%%%%%%%%%%%%%%%%%%%%%%%%%%%%%%%%%%%%%%
In this section, we first mention the objectives, and then present our proposed approach and algorithm to address the design problem.
\vspace{-0.1cm}
\subsection{Objectives}
We consider an optimization technique for the system illustrated in \figref{fig:diagram_CS} in order to determine the encoder and decoder mappings \textsf{E} and \textsf{D}, respectively, in the presence of channel error. More precisely, the aim of the VQ design is to find MSE-minimizing encoder regions $\{\mathcal{R}_i\}_{i=0}^{L-1}$ and decoder codebook $\mathcal{C} = \{\mathbf{c}_j\}_{j=0}^{L-1}$. On the other hand, a joint design of encoder and decoder cannot be implemented since the resulting optimization is analytically intractable. To address this issue, in \secref{subsec:opt enc}, we show how the encoding index $i \in \mathcal{I}$ (or equivalently encoder region $\mathcal{R}_i$) can be chosen to minimize the MSE for given values of codevectors $\{\mathbf{c}_j\}_{j=0}^{L-1}$. Then, in \secref{subsec:opt dec}, we derive an expression for the optimal decoder codebook (with respect to minimizing \eqref{eq:e2e dist}) for given encoder regions. Thereafter, in \secref{subsec:train}, we develop a VQ training design algorithm that combines the resulting necessary optimal rules.
\vspace{-0.3cm}
%%%%%%%%%%%%%%%%%%%%%%%%%%%%%%%%%%%%%%%%%%%%%%%%%%%%%%%%%%%%%%%%%%%%%%%%%
\subsection{Optimizing for Fixed Decoder Codebook} \label{subsec:opt enc}
%%%%%%%%%%%%%%%%%%%%%%%%%%%%%%%%%%%%%%%%%%%%%%%%%%%%%%%%%%%%%%%%%%%%%%%%%
\vspace{-0.1cm}
First, let us introduce the \textit{minimum mean-square error} (MMSE) estimator of the sparse input vector in the linear system model \eqref{eq:measurement} which is obtained as (see e.g. \cite[Chapter 11]{93:Kay})
\begin{equation} \label{eq:MMSE sparse}
    \tilde{\mathbf{x}}(\mathbf{y}) \triangleq  \mathbb{E} [\mathbf{X} | \mathbf{Y} = \mathbf{y}] \in \mathbb{R}^M. 
\end{equation}

Now, assume that the decoder codebook $\mathcal{C}=\{\mathbf{c}_j\}_{j=0}^{L-1}$ is known and fixed, then we focus on how the encoding index $i$ should be chosen to minimize the MSE given the observed measurement vector $\mathbf{y}=\mathbf{Ax + w}$. 

We rewrite the MSE in \eqref{eq:e2e dist} as
\begin{equation} \label{eq:rewrite MSE}
\begin{aligned}
    D &\triangleq \mathbb{E}[\|\mathbf{X - \widehat{X}}\|_2^2] = \mathbb{E}[\|\mathbf{X} - \mathbf{c}_J\|_2^2]& \\
    &\stackrel{(a)}{=} \! \int_{\mathbf{y}} \sum_{i \in \mathcal{I}} \textrm{Pr} \{I\!=\!i | \mathbf{Y\!=\!y}\} \mathbb{E} \left[\|\mathbf{X} \!-\! \mathbf{c}_J \|_2^2 | \mathbf{Y\!=\!y} , I\!=\!i\right] f (\mathbf{y}) d\mathbf{y}& \\
    &\stackrel{(b)}{=} \sum_{i \in \mathcal{I}} \int_{\mathbf{y}  \in \mathcal{R}_i} \left\{ \mathbb{E} \left[\|\mathbf{X} - \mathbf{c}_J \|_2^2 | \mathbf{Y=y} , I=i\right] \right\} f (\mathbf{y}) d\mathbf{y} ,&
\end{aligned}
\end{equation}
where $(a)$ follows from marginalization of the MSE over RV's $\mathbf{Y}$ and $I$. Further, $f(\mathbf{y})$ is the $N$-fold probability density function (p.d.f.) of the measurement vector. Also, $(b)$ follows by interchanging the integral and summation and the fact that $\textrm{Pr} \{I=i | \mathbf{Y=y}\} = 1$, $\forall \mathbf{y} \in \mathcal{R}_i$, and otherwise the probability is zero. Now, in order to minimize the MSE for given codevectors, it suffices to minimize the expression in the braces in the last expression of \eqref{eq:rewrite MSE} since $f(\mathbf{y})$ is positive. Then, the MSE-minimizing transmission index, denoted by $i^\star \in \mathcal{I}$, is given by
\begin{equation} \label{eq:proof enc}
\begin{aligned}
    i^\star &= \textrm{arg }\underset{i \in \mathcal{I}}{\textrm{min }} \mathbb{E} \left[\|\mathbf{X} - \mathbf{c}_J \|_2^2 | \mathbf{Y=y} , I=i\right]& \\
    &\stackrel{(a)}{=} \textrm{arg }\underset{i \in \mathcal{I}}{\textrm{min }} \mathbb{E}[\|\mathbf{c}_J\|_2^2  | \mathbf{Y\!=\!y}, I\!=\!i] \!-\! 2\mathbb{E}[\mathbf{X}^T \mathbf{c}_J  | \mathbf{Y\!=\!y}, I\!=\!i]& \\
    &\stackrel{(b)}{=} \textrm{arg }\underset{i \in \mathcal{I}}{\textrm{min }} \mathbb{E}[\|\mathbf{c}_J\|_2^2 \big | I=i] - 2\mathbb{E}[\mathbf{X}^T \big | \mathbf{Y=y}] \mathbb{E}[\mathbf{c}_J \big | I=i],&
\end{aligned}
\end{equation}
where $(a)$ follows from the fact that $\mathbf{X}$ is independent of $I$, conditioned on $\mathbf{Y}$, hence, $\mathbb{E} \left[\|\mathbf{X}\|_2^2 | \mathbf{Y\!=\!y}, I\!=\!i \right] = \mathbb{E} \left[\|\mathbf{X}\|_2^2 | \mathbf{Y\!=\!y}\right]$ which is pulled out of the optimization. $(b)$ follows from Markov chains $\mathbf{Y} \rightarrow I \rightarrow \mathbf{c}_J$ and $\mathbf{X} \rightarrow \mathbf{Y} \rightarrow I$. Next, note that introducing channel transition probabilities \eqref{eq:channel trans} and the MMSE estimator \eqref{eq:MMSE sparse}, the last equality in \eqref{eq:proof enc} can be expressed as
\begin{equation} \label{eq:final enc}
    i^\star = \textrm{arg }\underset{i \in \mathcal{I}}{\textrm{min}} \left\{ \sum_{j=0}^{L-1} P(j|i) \left\| \mathbf{c}_j \right\|_2^2
    - 2 \tilde{\mathbf{x}}(\mathbf{y})^T \sum_{j=0}^{L-1} P(j|i) \mathbf{c}_j \right\} .
\end{equation}

It is also straightforward to show that in the case of error-free channel ($P(j|i) = 0$, $\forall i \neq j \in \mathcal{I}$), given the observations $\mathbf{y}$ and a noiseless channel, the quantizer encoder that minimizes the MSE (assuming known codevectors) is given by
\begin{equation} \label{eq:optimal enc}
    i^\star = \textrm{arg }\underset{i \in \mathcal{I}}{\textrm{min}} \left\{ \left\| \mathbf{c}_i \right\|_2^2
    - 2 \tilde{\mathbf{x}}(\mathbf{y})^{T} \mathbf{c}_i \right\}.
\end{equation}
\vspace{-0.75cm}
%%%%%%%%%%%%%%%%%%%%%%%%%%%%%%%%%%%%%%%%%%%%%%%%%%%%%%%%%%%%%%%%%%%%%%%%%
\subsection{Optimizing for Fixed Encoding Regions} \label{subsec:opt dec}
%%%%%%%%%%%%%%%%%%%%%%%%%%%%%%%%%%%%%%%%%%%%%%%%%%%%%%%%%%%%%%%%%%%%%%%%%
\vspace{-0.1cm}

Adopting the MSE criterion, it is straightforward to show the codevectors which minimize $D$ in \eqref{eq:e2e dist} are obtained  by letting $\mathbf{c}_j$ represent the MMSE estimator of the vector $\mathbf{X}$ based on the received index $j$, that is,  $\mathbf{c}_j$ should be chosen as
\begin{equation} \label{eq:opt dec}
    \mathbf{c}_j^\star =  \mathbb{E}[\mathbf{X} | J=j] , \hspace{0.15cm} j \in \mathcal{I}.
\end{equation}
Now, let define
\begin{equation} \label{eq:Pi Pj}
\begin{aligned}
    &P(i) \! \triangleq \! \textrm{Pr}(I\!=\!i) \!=\! \textrm{Pr}(\mathbf{Y} \!\in\! \mathcal{R}_i),
    P(j) \!\triangleq \!\textrm{Pr}(J\!=\!j) \!=\! \sum_{i} P(j|i) P(i),& \\
    &\hspace{1.25cm} P(i|j) \triangleq  \textrm{Pr}(I = i | J = j) = P(j|i)P(i)/P(j),&
\end{aligned}
\end{equation}
then, the expression for $\mathbf{c}_j^\star$ can be rewritten as
\begin{equation} \label{eq:opt dec final}
\begin{aligned}
    &\mathbf{c}_j^\star =  \mathbb{E}[\mathbf{X} | J=j]
    = \sum_{i} P(i | j) \mathbb{E}[\mathbf{X} | J=j,I=i]&\\
    &\!\stackrel{(a)}{=} \! \sum_i P(i | j) \mathbb{E}[\mathbf{X} | I\!=\!i]
    \!\stackrel{(b)}{=} \! \frac{\sum_i P(j|i) P(i) \int_{\mathbf{y}}\mathbb{E}[\mathbf{X}|\mathbf{Y\!=\!y}] f(\mathbf{y}|i) d\mathbf{y}}{\sum_i P(j|i) P(i)}& \\
    &\stackrel{(c)}{=} \frac{\sum_i P(j|i) \int_{\mathcal{R}_i} \tilde{\mathbf{x}}(\mathbf{y}) f(\mathbf{y}) d\mathbf{y} }{\sum_i P(j|i) \int_{\mathcal{R}_i} f(\mathbf{y}) d\mathbf{y}},&
\end{aligned}
\end{equation}
where $(a)$ follows from the Markov chain $\mathbf{X} \rightarrow I \rightarrow J$, and $(b)$ from marginalization over $\mathbf{Y}$, the Markov chain $\mathbf{X} \rightarrow \mathbf{Y} \rightarrow I$, and \eqref{eq:Pi Pj} which translates $(a)$ to the known parameters. Moreover, $f(\mathbf{y}|i)$ is the conditional p.d.f. of $\mathbf{Y}$ given that $\mathbf{Y} \in \mathcal{R}_i$. Also, $(c)$ is followed by \eqref{eq:MMSE sparse} and by the fact that $f(\mathbf{y}|i)=0$ for $\mathbf{y} \notin \mathcal{R}_i$.

As a special case where the channel is error-free (i.e., $P(j|i) = 0$, $\forall j \neq i \in \mathcal{I}$), given the known encoding index, the MSE-minimizing codebook, denoted by $\mathcal{C^\star} = \{\mathbf{c}_i^\star\}_{i=0}^{L-1}$, must satisfy
\begin{equation} \label{eq:optimal dec noiseless}
    \mathbf{c}_i^\star =  \mathbb{E}[\mathbf{X} | I=i].
\end{equation}

In general, it is not easy to derive closed form solutions for the necessary optimality conditions \eqref{eq:final enc} and \eqref{eq:opt dec final}. However, in practice, since the alphabets of the set $\mathcal{I}$ are finite, these expressions can be computed using a database. In the next section, we use these principles in order to design a training algorithm for the CS-VQ system.

\vspace{-0.4cm}
%%%%%%%%%%%%%%%%%%%%%%%%%%%%%%%%%%%%%%%%%%%%%%%%%%%%%%%%%%%%%%%%%%%%%%%%%%%%%%%%
\subsection{Training Design Algorithm} \label{subsec:train}
%%%%%%%%%%%%%%%%%%%%%%%%%%%%%%%%%%%%%%%%%%%%%%%%%%%%%%%%%%%%%%%%%%%%%%%%%%%%%%%%
\vspace{-0.1cm}

The results presented in \secref{subsec:opt enc} and \secref{subsec:opt dec} can be utilized to formulate an iterate-alternate training algorithm for the problem of interest. Similar to the %classic \textit{Lloyd} algorithm for quantizer design in the case of error-free channels \cite{91:Gersho}, and
\textit{generalized Lloyd} algorithm for noisy channels \cite{90:Farvardin}, we propose a VQ training method for the design problem in this work which is summarized in \algref{alg:Lloyd}.

\vspace{-0.2cm}
\begin{algorithm}
\caption{: Channel-optimized vector quantization training algorithm for compressed sensing measurements}\label{alg:Lloyd}
\begin{algorithmic}[1]
\STATE{\textbf{input:} measurement vector: $\mathbf{y}$, channel probabilities: $P(j|i)$, bit budget: $B$ bit/dimesnion.}
\STATE{\textbf{compute:} $\tilde{\mathbf{x}}(\mathbf{y})$ in \eqref{eq:MMSE sparse}.}
\STATE{\textbf{initialize: } $\mathcal{C} = \{\mathbf{c}_j\}_{j=0}^{L-1}$ where $L=2^B$}
\REPEAT
    \STATE{Update and fix the codevectors, then calculate the encoding indexes using \eqref{eq:final enc}.}
    \STATE{Update and fix the encoding indexes, then calculate the codevectors using \eqref{eq:opt dec}.}
\UNTIL{convergence}
\STATE{\textbf{output: } $\{\mathcal{R}_i\}_{i=0}^{L-1}$ , $\mathcal{C}=\{\mathbf{c}_j\}_{j=0}^{L-1}$ }
\end{algorithmic}
\end{algorithm}
\vspace{-0.2cm}

The following remarks can be considered for implementing \algref{alg:Lloyd}. In step (1), besides the channel transition probabilities $P(j|i)$, we assume that the statistics of the sparse source are given. % as in \eqref{eq:pdf X}.
In step (2), computing the optimal estimator $\tilde{\mathbf{x}}(\mathbf{y})$ in \eqref{eq:MMSE sparse} is not practically feasible since the size of support set $\mathcal{S}$ increases exponentially with size of the sparse source $\mathbf{X}$ (cf. \cite{09:Elad}). Instead, we will approximate $\tilde{\mathbf{x}}(\mathbf{y})$ using the output of the low-complexity greedy \textit{orthogonal matching pursuit} (OMP) reconstruction algorithm \cite{07:Tropp,08:Blumensath}. In step (3), the codevectors can be initialized randomly. Convergence in step (7) may be checked by tracking the MSE, and terminate the iterations when the relative improvement is small enough. In principle (and ignoring issues such as numerical precision), the iterative design in \algref{alg:Lloyd} always converges to a local optimum since when the criteria in steps (5) and (6) of the algorithm are invoked, the performance can only leave unchanged or improved, given the updated indexes and codevectors. This is a common rationale behind the proof of convergence for such iterative algorithms (see e.g. \cite[Lemma 11.3.1]{91:Gersho}). However, nothing can be generally guaranteed about the global optimality of this algorithm.
\vspace{-0.4cm}
%%%%%%%%%%%%%%%%%%%%%%%%%%%%%%%%%%%%%%%%%%%%%%%%%%%%%%%%%%%%%%%%%%%%%%%%%
\section{Experiments and Results} \label{sec:numerical}
%%%%%%%%%%%%%%%%%%%%%%%%%%%%%%%%%%%%%%%%%%%%%%%%%%%%%%%%%%%%%%%%%%%%%%%%%
\vspace{-0.2cm}

%%%%%%%%%%%%%%%%%%%%%%%%%%%%%%%%%%%%%%%%%%%%%%%%%%%%%%%%%%%%%%%%%%%%%%%
\subsection{Related Quantization Methods} \label{subsec:other schemes}
%%%%%%%%%%%%%%%%%%%%%%%%%%%%%%%%%%%%%%%%%%%%%%%%%%%%%%%%%%%%%%%%%%%%%%%
\vspace{-0.1cm}
In order to study the effect of inherent sparsity of the source and channel errors on the VQ design for CS measurements, we introduce two quantization schemes:

\begin{enumerate}
\item For a noisy channel, an alternative quantization is the use of COVQ designed for the input vector $\mathbf{Y}$ (see \cite{90:Farvardin} for details) aiming to minimize the \textit{average quantization distortion} when sending $\mathbf{Y}$ over the channel. The design strategy of the encodeing/decoding rules is as follows: for a quantization rate $B$ bits/vector, and for a fixed codebook $\mathcal{G} \!\triangleq\! \{\mathbf{g}_{j} \! \in \! \mathbb{R}^N\}_{j=0}^{L-1}$, where $L \!=\! 2^B$, and known channel transition probability $P(j|i)$, a transmission index $i^\sharp \! \in \! \mathcal{I}$ is chosen as
    \begin{equation} \label{eq:CO region}
        %\begin{aligned}
            i^\sharp = \textrm{arg }\underset{i \in \mathcal{I}}{\textrm{min}}   \sum_{j=0}^{L-1} \|\mathbf{y} \!-\! \mathbf{g}_{j}\|_2^2 P(j|i).
        %\end{aligned}
        \end{equation}
    Now, for the given index characterized by \eqref{eq:CO region} and channel transition probability $P (j|i)$, the MMSE reproduction codevectors (with respect to minimizing average quantization distortion) satisfy
    \begin{equation} \label{eq:CO cent}
            \mathbf{g}_{j}^\sharp = \mathbb{E}[\mathbf{Y} | J=j]  , \hspace{0.5cm}  j \in \mathcal{I}.
    \end{equation}
    Similar to \algref{alg:Lloyd}, this procedure is alternated between \eqref{eq:CO region} and \eqref{eq:CO cent} and then iterated which converges to (locally) optimum codevectors and indexes. Finally, a (suboptimal) CS recovery algorithm (here OMP) takes the codevector as an input and reconstruct the sparse source. As this scheme exploit channel-optimized rules for minimizing only quantization distortion, we will label this method as \textit{``COVQ-Q''} in our experiments.

\item To study the effect of channel errors on the design, we exploit channel-unoptimized VQ (CUVQ) which is indeed the case where knowledge about channel condition is not provided at the transmitter and receiver. For this scheme, the encoder index and decoder codevectors are obtained by \eqref{eq:optimal enc} and \eqref{eq:optimal dec noiseless}, respectively, in an iterate-alternate procedure similar to \algref{alg:Lloyd}. We will refer to this method as \textit{``CUVQ-E2E''} since a designer is not aware of channel imperfections, but the end-to-end distortion is considered.
\end{enumerate}

%%%%%%%%%%%%%%%%%%%%%%%%%%%%%%%%%%%%%%%%%%%%%%%%%%%%%%%%%%%%%%%%%%%%%%%%%%%%
\subsection{Experimental Setups and Results} \label{subsec:setups}
%%%%%%%%%%%%%%%%%%%%%%%%%%%%%%%%%%%%%%%%%%%%%%%%%%%%%%%%%%%%%%%%%%%%%%%%%%%%
\vspace{-0.1cm}
We examine the performance of the proposed algorithm using normalized MSE (NMSE) defined as
%\begin{equation} \label{eq:NMSE}
    $\textrm{NMSE} \triangleq \frac{1}{K} \mathbb{E}[\|\mathbf{X}-\mathbf{\widehat{X}}\|_2^2]$.
%\end{equation}
To measure the level of under-sampling, we define the measurement rate as
%\begin{equation} \label{eq:fom}
    $\alpha \triangleq N/M$.
%\end{equation}
We generate the Gaussian sparse source $\mathbf{X}$, and the sensing matrix model $\mathbf{\Phi}$ as described in \secref{subsec:CS}. In order to focus on distortion due to quantization and channel, we assume that the additive measurement noise $\mathbf{W}$ in our linear model \eqref{eq:measurement} is negligible. Further, $\tilde{\mathbf{x}}(\mathbf{y})$ in \eqref{eq:MMSE sparse} is (approximately) calculated by the low-complexity OMP algorithm \cite{07:Tropp,08:Blumensath}. We assume a binary symmetric channel (BSC) with bit cross-over probability $\epsilon$ specified by
\begin{equation} \label{eq:BSC}
          P(j | i) = \epsilon^{H_B(i,j)} (1 - \epsilon)^{B - H_B(i,j)},
\end{equation}
where $0 \leq \epsilon \leq 1/2$ represents bit cross-over probability (assumed known), and $H_B(i,j)$ denotes the Hamming distance between $B$-bit binary codewords representing the channel input and output indexes $i$ and $j$. To evaluate the NMSE, we run \algref{alg:Lloyd} for $10^5$ trials and average over the number of simulation rounds. We refer to our proposed design as \textit{``COVQ-E2E''} in the following results.

\begin{figure}
  \begin{center}
  \includegraphics[width=\columnwidth,height=6.5cm]{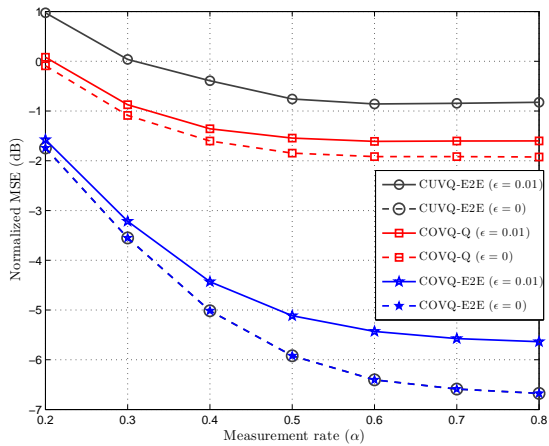}\\
  \vspace{-0.3cm}
  \caption{NMSE vs. measurement rate $\alpha \!=\! N/M$.}   \label{fig:MSE_FOM}
  \vspace{-0.75cm}
  \end{center}
\end{figure}

Using parameters of source vector size $M=20$, sparsity level $K=2$ (sparsity ratio $= 10\%$) and quantization rate $B=8$ bits/dimension, we show performance comparison (in terms of NMSE) for quantizer design methods in \figref{fig:MSE_FOM} as a function of measurement rate $\alpha$, and the curves correspond to channel bit cross-over probabilities $\epsilon = 0$ and $\epsilon = 0.01$. It can be seen as the number of measurements increases, the performance improves due to the fact that more measurements yield a more precise estimate of the source at a fixed quantization rate $B$ and cross-over probability $\epsilon$. At $\epsilon=0$, the COVQ-E2E and CUVQ-E2E designs coincide as expected from theory. However, the COVQ-Q design does not take into account the end-to-end distortion which yields to a poor performance at low channel noise. At $\epsilon = 0.01$, not surprisingly, the COVQ-E2E outperforms the other ones. It is also revealed the CUVQ-E2E design performs inferior to the COVQ-Q since it does not consider channel transition probabilities. Using the proposed COVQ-E2E design, we gain notable $4$ dB improvement at $\alpha\!=\!0.5$.

In order to investigate the effect of channel noise on the performance at fixed measurement and quantization rates, we compare the performance of the different quantization designs in Table 1 for various channel bit cross-over probabilities $\epsilon$. As one would expect, the COVQ-E2E design gives the best performance. Further, comparing all methods together, it can be observed that the loss in the performance of the CUVQ-E2E design grows substantially as the channel becomes noisier. Thus, it can be concluded that the channel condition knowledge is crucial throughout the design procedure.

\begin{table} \label{table: MSE_epsilon}
    \caption{Normalized MSE (in dB) vs. bit cross over probability $\epsilon$ for $M\!=\!20$, $K\!=\!2$, $B\!=\!8$ bit/dimesnion and $\alpha \!=\! 0.5$.}
    %\vspace{-0.1cm}
    \begin{center}
    \begin{tabular}{ | l | c | c | c |}
    \hline
    $\epsilon$ & 0 & 0.001 & 0.005 \\ \hline
    COVQ-E2E & -5.9371 & -5.7616 & -5.3308 \\ \hline
    COVQ-Q & -1.8357 & -1.8262 & -1.6737  \\ \hline
    CUVQ-E2E & -5.9371 & -4.7308 & -2.7242 \\
    \hline \hline
    $\epsilon$ & 0.01 & 0.05 & 0.1\\ \hline
    COVQ-E2E & -4.8947 & -3.0293 & -1.8257 \\ \hline
    COVQ-Q & -1.5370 & -0.8508 & -0.4511 \\ \hline
    CUVQ-E2E & -0.4334 & 4.7124 &  6.7608 \\ \hline
    \end{tabular}
    \vspace{-0.25cm}
    \end{center}
\end{table}

\begin{figure}
%  \vspace{-0.5cm}
  \begin{center}
  \includegraphics[width= \columnwidth,height=6.5cm]{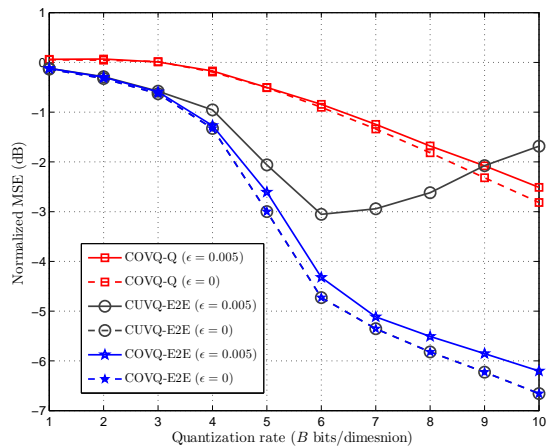}\\
  \vspace{-0.3cm}
  \caption{NMSE vs. quantization rate $B$ (bits/vector).}
  \label{fig:MSE_rate}
  \vspace{-0.75cm}
  \end{center}
\end{figure}

Next, we examine what the behavior of the performance would be if we increase quantization rate. The parameters are chosen as $M\!=\!20$, $K\!=\!2$ and $\alpha \!=\! 0.5$, and we vary the quantization rate. In \figref{fig:MSE_rate}, we illustrate the performance of the design approaches for bit cross-over probabilities $\epsilon=0$ and $\epsilon = 0.005$. As expected, the performance of the COVQ-E2E design is superior to that of the COVQ-Q and CUVQ-E2E designs, where we can gain more than $3$ dB improvement at high quantization rates. Not surprisingly, the performance of the COVQ-E2E and CUVQ-E2E designs are the same for an error-free channel. Further, the CUVQ-E2E design shows a different behavior by increasing $B$ at $\epsilon = 0.005$. This is due to the fact that at a fixed $\epsilon$, although the quantization distortion decreases as quantization rate increases, the probability of correct index reception decreases exponentially. Hence, the performance of the CUVQ-E2E design, which is unaware of channel condition, considerably degrades at high rates.

%\vspace{-0.35cm}
\section{Conclusions} \label{sec:conclusion}
%\vspace{-0.2cm}
In literature, the problem of VQ design for CS measurements has been addressed without considering channel noise. To the best of our knowledge, this is the first time that channel imperfections have been taken into account through the design of COVQ for CS measurements. Adopting the end-to-end MSE criterion, we derive necessary optimal encoding and decoding principles, and utilize these conditions to develop an algorithm for training a COVQ for encoding CS measurements and reconstructing a sparse source. Numerical results have shown the promising performance gained by using our proposed optimal design. 

\bibliographystyle{IEEEtran}
\bibliography{IEEEfull,bibliokthPasha}
\end{document}